\documentclass[11pt]{article}
\usepackage[sort&compress,square,comma,numbers]{natbib}

\usepackage[labelfont=bf]{caption} 
\usepackage[font={small}]{caption}   

\usepackage{subcaption}
\usepackage{amsmath}
\usepackage{amssymb}
\usepackage{url}
\usepackage{placeins}
\usepackage{textcomp}
\usepackage{graphicx}
\usepackage{float} 
\usepackage{chngpage}
\usepackage{epstopdf}
\usepackage[usenames,dvipsnames]{color}    

\usepackage[usenames,dvipsnames]{color}    
\usepackage{soul}   
\setstcolor{red} 
\setul{}{1.5pt}  

\title{\textbf{Direction reconfigurable non-reciprocal acousto-optic modulator on chip}}
\author{Donggyu B Sohn, Gaurav Bahl$^{\ast}$\\
	\\
	\footnotesize{Mechanical Science and Engineering, University of Illinois at Urbana-Champaign,}\\
	\footnotesize{Urbana, Illinois, USA}\\
	\footnotesize{$^\ast$To whom correspondence should be addressed; E-mail: bahl@illinois.edu}
	}

\date{}

\begin{document}
\maketitle

\begin{abstract}
Non-reciprocal components are essential in photonic systems for protecting light sources and for signal routing functions.
Acousto-optic methods to produce non-reciprocal devices offer a foundry-compatible alternative to magneto-optic solutions and are especially important for photonic integration.
In this paper, we experimentally demonstrate a dynamically reconfigurable non-reciprocal acousto-optic modulator at telecom wavelength.
The modulator can be arranged in a multitude of reciprocal and non-reciprocal configurations by means of an external RF input.
The dynamic reconfigurability of the device is enabled by a new cross-finger interdigitated piezoelectric transducer that can change the directionality of the reciprocity-breaking acoustic excitation based on the phase of the RF input. 
The methodology we demonstrate here may enable new avenues for direction dependent signal processing and optical isolation.
\end{abstract}

	\section*{Introduction}

	Optical isolators and circulators are crucial components in photonic circuits for protecting light sources from backscattering and for achieving direction dependent signal routing \cite{Hogan:1952,aplet1964,jalas2013}. 
	These functions can be only achieved by components that are non-reciprocal, i.e. that have asymmetric scattering matrices that imply different port-to-port propagation in opposing directions.
	The traditional approach to produce non-reciprocal components employs the Faraday rotation effect in magneto-optic media, and has recently been implemented for chip-scale photonics through heterogeneous bonding of ferrites \cite{Ross:11,Huang:17,Zhang:19}. 
	However, this approach is not yet feasible in foundries due to materials limitations, and challenges with the generation and confinement of magnetic fields on-chip.
	As a result, while foundry-based integrated photonic systems have made tremendous progress, developments in foundry-compatible integrated non-reciprocal devices have been relatively slow.


	An alternative approach is to use spatio-temporal modulation or momentum-biasing of the medium to produce magnetless non-reciprocity. This class of methods, of which specific approaches employ synthetic magnetism \cite{Fang:12,Tzuang2014,Fang:2017}, inter-band photonic transition \cite{Hwang:97,Yu2009,KangM.2011,lira2012,Kim2015,Dong2015,Kim2016,Sohn18,Kittlaus2018,Peterson:18}, angular momentum biasing~\cite{Sounas:13,Sounas:14,Fleury:16}, and phase modulation \cite{Fujita:00,Doerr:11,Shen:16, Ruesink:16,Peterson:17, Bernier17} is particularly attractive since they leverage common dielectrics and can produce linear non-reciprocal responses. Moreover, since these methods employ optical and electrical modulations, the non-reciprocity can be activated and deactivated when needed.
	

	In this context, we previously demonstrated an on-chip non-reciprocal modulator \cite{Sohn18} that uses a traveling acoustic wave to produce indirect inter-band transitions in a photonic resonator. The photoelastic perturbations of the dielectric material caused by the acoustic wave possess high momentum and low frequency, which is essential for the momentum biasing technique \cite{KangM.2011,Dong2015,Kim2015,Kim2016,Sohn18,Kittlaus2018}. The underlying phase matching requirement only permits this optical coupling in one direction, depending on the propagation momentum of the acoustic wave, and hence produces a non-reciprocal modulation response.
	Since the device geometry, e.g. the interdigitated transducer that actuates the acoustic wave, cannot be changed once fabricated, the direction of non-reciprocity is fixed.
	In this work, we introduce a new design for on-chip non-reciprocal modulators that permits dynamic direction reconfigurability of the non-reciprocal effect. With the approach demonstrated here, the level of transparency (in terms of whether or not the signal is modulated) or opacity in the forward and backward directions can be set independently.
	%

	\section*{Theory}

	\begin{figure}[tb]
	\begin{adjustwidth}{-1in}{-1in}
		\makebox[\textwidth][c]{\includegraphics[width=1.3\textwidth]{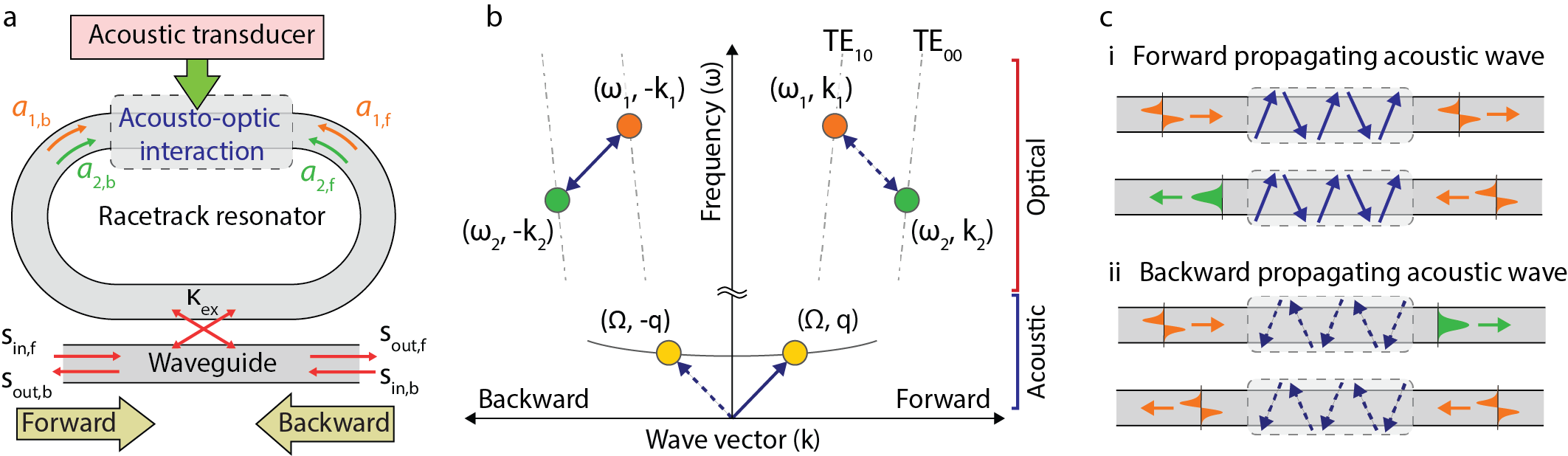}}
		\centering
		\caption{\textbf{Conceptual schematic of the direction reconfigurable non-reciprocal acousto-optic modulator.} \textbf{(a)} The device is composed of a racetrack resonator supporting two optical modes (TE$_{00}$ and TE$_{10}$) that is evanescently coupled to a single mode waveguide. The inter-mode acousto-optic scattering is enabled over the linear part of the racetrack resonator.
		\textbf{(b)} Phase-matching requirements between the optical fields and acoustic excitation represented in frequency momentum space enable the non-reciprocal modulation. Here we illustrate a case where the resonance frequency of the TE$_{10}$ ($\omega_1$,$k_1$) mode is higher than that of the TE$_{00}$ ($\omega_2$,$k_2$) mode. The existence of a forward (backward) propagating acoustic wave at ($\Omega$,$q$) satisfies the phase matching condition for only the backward (forward) circulating optical modes.
		\textbf{(c)} Illustration of the acousto-optic interaction in the linear region of the racetrack, with (i) forward and (ii) backward propagating acoustic waves. When the light and the acoustic wave counter-propagate, the inter-mode photonic transition can occur. However, when the light and the acoustic wave co-propagate the scattering is suppressed since the phase matching condition is not satisfied.
		}
		\label{Recon1}
	\end{adjustwidth}
\end{figure}

	A non-reciprocal modulator can be obtained through intermodal scattering between two optical modes of the photonic system that are distinct in both frequency and momentum space \cite{Sohn18,Kittlaus2018,Yu2009}.
	 In this work, we employ an optical racetrack resonator that supports two (four if we count opposite direction partners) distinct circulating optical modes -- the quasi-TE$_{00}$ ($\omega_1$, $\pm k_1$) and quasi-TE$_{10}$ ($\omega_2$, $\pm k_2$) modes as shown in Fig.~\ref{Recon1}a. For compactness of notation, we drop the `quasi' prefix in further description. 
	The principle of intermodal scattering in this resonator is illustrated in the $\omega-k$ diagram in Fig.~\ref{Recon1}b.
	We consider the specific case where the resonance frequency of the TE$_{10}$ mode is higher than that of the TE$_{00}$ mode (Fig.~\ref{Recon1}b).
	An acoustic wave with frequency and momentum ($\Omega$, $q$) can photoelastically modulate the resonator  to enable intermodal scattering between the optical modes. This scattering of course requires that the phase matching condition $\Omega \, = \, \omega_1 \, - \, \omega_2$ and $q \, = \, k_1 \, - \, k_2$ is satisfied.
	In example here, if the acoustic wave propagates in the forward direction (i.e. with positive wave vector), only the backward propagating (i.e. negative wave vector) optical modes are coupled as illustrated in Fig.~\ref{Recon1}c(i). This coupling allows the mode conversion from the TE$_{00}$ mode to the TE$_{10}$ mode through anti-Stokes scattering or from the TE$_{10}$ mode to TE$_{00}$ mode through Stokes scattering. 
	However, this same acoustic wave does not satisfy the phase matching condition for the forward propagating optical modes due to momentum mismatch.
	On the other hand, if the acoustic wave direction is reversed, i.e. the acoustic wave propagates in the backward direction (Fig.~\ref{Recon1}c(ii)), only the forward propagating optical modes would be coupled but the backwards propagating optical modes remain unaffected. 
	Thus, the direction of the non-reciprocal scattering can in principle be controlled by dynamically adjusting the direction of the acoustic wave.

	
	In our previous study on this topic \cite{Sohn18}, we employed an interdigitated transducer (IDT) patterned on piezoelectric film to launch the acoustic wave. 
	IDTs are typically structured as a periodic array of paired electrodes, each carrying one polarity of an RF drive signal. This RF electrical stimulus mechanically deforms the material due to the piezoelectric effect, and launches acoustic waves that match the acoustic dispersion relation, with the wave vector magnitude $|q|$ being set through the physical periodicity of the IDT. When stimulated, the IDT necessarily launches acoustic waves outwards with both $+q$ and $-q$ wave vectors in opposite directions. Depending on the physical placement of the photonic device in relation to the IDT, only one of these wave vectors may be utilized~\cite{Tadesse2014,Balram2016,Sohn18,Liu:19}, or a standing wave may be utilized~\cite{Ghosh:15,Shao19}, either of which cannot be changed once fabricated.
	Moreover, while unidirectional IDT designs (i.e. only $+q$ or only $-q$) have been developed previously~\cite{Lu:2019aa}, these IDT variants also cannot be reconfigured once fabricated.
	Since our goal in this work is to be able to dynamically reconfigure the non-reciprocity without having to make physical changes to the structure, a new transducer design is needed.
	
	\vspace{12pt}
	
	Here, we explore a different approach to synthesize an acoustic wave with dynamically reconfigurable momentum through the superposition of two standing acoustic waves with relative spatial and temporal phase offsets. 
	To understand this approach let us consider two frequency degenerate co-axial standing waves having a quarter wavelength relative shift in space, and define the $z$ as the axis on which these standing waves are formed. We can represent these two waves mathematically as $u_1 = \psi(x,y) \cos(qz)\cos(\Omega t)$ and $u_2 = \psi(x,y)\sin(qz)\cos(\Omega t+\theta_t)$ with $\theta_t$ being a phase that we can assign based on driving. Here, $\psi(x,y)$ is the cross-sectional mode shape of the acoustic wave including its amplitude, and $\theta_t$ represents the relative temporal phase of the acoustic wave $u_2$ with respect to the acoustic wave $u_1$. The superposition of these two standing waves produces the acoustic excitation: 
	\begin{align}
	\label{eqn:superpositionequation}
	u & = u_1 + u_2 \\
		& = \frac{1}{2}\psi(x,y) \left\lbrace (1+e^{i(\theta_t-\pi/2)})e^{i(qz-\Omega t)}+(1+e^{i(\theta_t+\pi/2)})e^{i(-qz-\Omega t)} \right\rbrace  + c.c.  \notag
	\end{align}
	where the first term on the RHS is a forward propagating acoustic wave ($+q$ wave vector) and the second term is a backward propagating acoustic wave ($-q$ wave vector). The most interesting cases appear when the relative temporal phase is either set to $\theta_t\,=\,\pi/2$ or $\theta_t\,=\,3\pi/2$, at which point the excitation becomes a pure traveling wave. 
	For $\theta_t\,=\,\pi/2$, the synthesized acoustic wave propagates only in the backward direction, and Eqn.~\ref{eqn:superpositionequation} simplifies to:
	\begin{align}
		\left. u \right|_{\theta_t=\pi/2} =  \psi(x,y) \cos(-qz-\Omega t)
	\end{align}
    Similarly, For $\theta_t\,=\,3\pi/2$, the synthesized acoustic wave propagates in the forward direction, and Eqn.~\ref{eqn:superpositionequation} simplifies to:
	\begin{align}
		\left. u \right|_{\theta_t=3\pi/2} = \psi(x,y) \cos(qz-\Omega t)
	\end{align}
	This means that by simply adjusting the relative temporal phase, we can dynamically reconfigure the fraction of forward or backward propagating wave contributions , and therefore dynamically reconfigure a non-reciprocal modulator or optical isolator based on this platform.

	\vspace{12pt}

	
	We now analyze the acousto-optic intermodal interaction with the above synthetized traveling acoustic wave using the coupled equations of motion: 
	\begin{align}
		\dfrac{\partial}{\partial t} 
		A_m
		=
		(-iW-\Gamma-G_m)A_m+Ks_{in,m}e^{-i\omega_l t}.
	\end{align}
	where the subscript $m$ can represent either forward ($f$) or backward ($b$) propagation, and indicates a defined directionality of light propagating through the system. Other system parameters are represented by matrices
	\begin{align*}
		A_m=
		\begin{pmatrix}
		a_{1,m}\\
		a_{2,m} 
		\end{pmatrix},
		~~~	
		W=
		\begin{pmatrix}
		\omega_1& 0 \\
		0 & \omega_2 
		\end{pmatrix},
		~~~
		\Gamma=
		\begin{pmatrix}
		\kappa_1& 0 \\
		0 & \kappa_2 
		\end{pmatrix},
		~~~		
		K=
		\begin{pmatrix}
		\sqrt{\kappa_{ex1}}\\
		\sqrt{\kappa_{ex2}} 
		\end{pmatrix}.
	\end{align*}
	 Here, $a_{1,m}$ and $a_{2,m}$ are the intracavity field amplitudes within the TE$_{10}$ and TE$_{00}$ modes respectively, $\omega_{1}$ and $\omega_{2}$ are the corresponding resonance frequencies, $\kappa_{ex1}$ and $\kappa_{ex2}$ are the corresponding external coupling rates between the waveguide and the resonator modes, and $\kappa_1$ and $\kappa_2$ are the corresponding loaded loss rates. The variable $\omega_l$ represents the input laser frequency and $s_{in,m}$ is the input field to the waveguide. Most of variables of interest are illustrated in Fig.~\ref{Recon1}a. The output field amplitude $s_{out}$ can then be expressed as:
	\begin{align}
	s_{out,m}=s_{in,m}-K^TA_m.
	\label{outputeq}
	\end{align}	
	Due to the phase matching condition, the coupling between the forward propagating optical modes is only induced by the backward propagating component of the acoustic excitation (see Fig.~\ref{Recon1}b). Thus, the corresponding coupling term becomes:
	\begin{align}
		G_f=
		\begin{pmatrix}
		0 & g(1+e^{i(\theta_t+\pi/2)})e^{i\Omega t} \\
		g(1+e^{-i(\theta_t+\pi/2)})e^{-i\Omega t} & 0
		\end{pmatrix}.
	\end{align}
	where $g$ is the optomechanical coupling coefficient proportional to the cross-sectional overlap integral of the two optical and acoustic waves, which will discuss in detail later. Similarly, the coupling between the backward propagating optical modes is only induced by the forward propagating component of the acoustic excitation. Thus, the corresponding coupling term becomes:
	\begin{align}
		G_b=
		\begin{pmatrix}
		0 &  g(1+e^{i(\theta_t-\pi/2)})e^{i\Omega t} \\
		 g(1+e^{-i(\theta_t-\pi/2)})e^{-i\Omega t} & 0
		\end{pmatrix}.
	\end{align}	
	As shown in the equations above, the intermodal coupling $G_m$ between the optical modes is always a function of the relative temporal phase $\theta_t$. Specifically, when the phase is $\theta_t~=~\pi/2$ only the forward propagating light experiences the intermodal scattering since the backward coupling term $G_b=0$. Similarly, when the phase is $\theta_t~=~3\pi/2$ only the backward propagating light experiences the intermodal scattering, since $G_f=0$.

	\vspace{24pt}

	\section*{Cross-finger interdigitated transducer (CFIDT)}

	\begin{figure}[t!]
	\begin{adjustwidth}{-1in}{-1in}
		\makebox[\textwidth][c]{\includegraphics[width=1.3\textwidth]{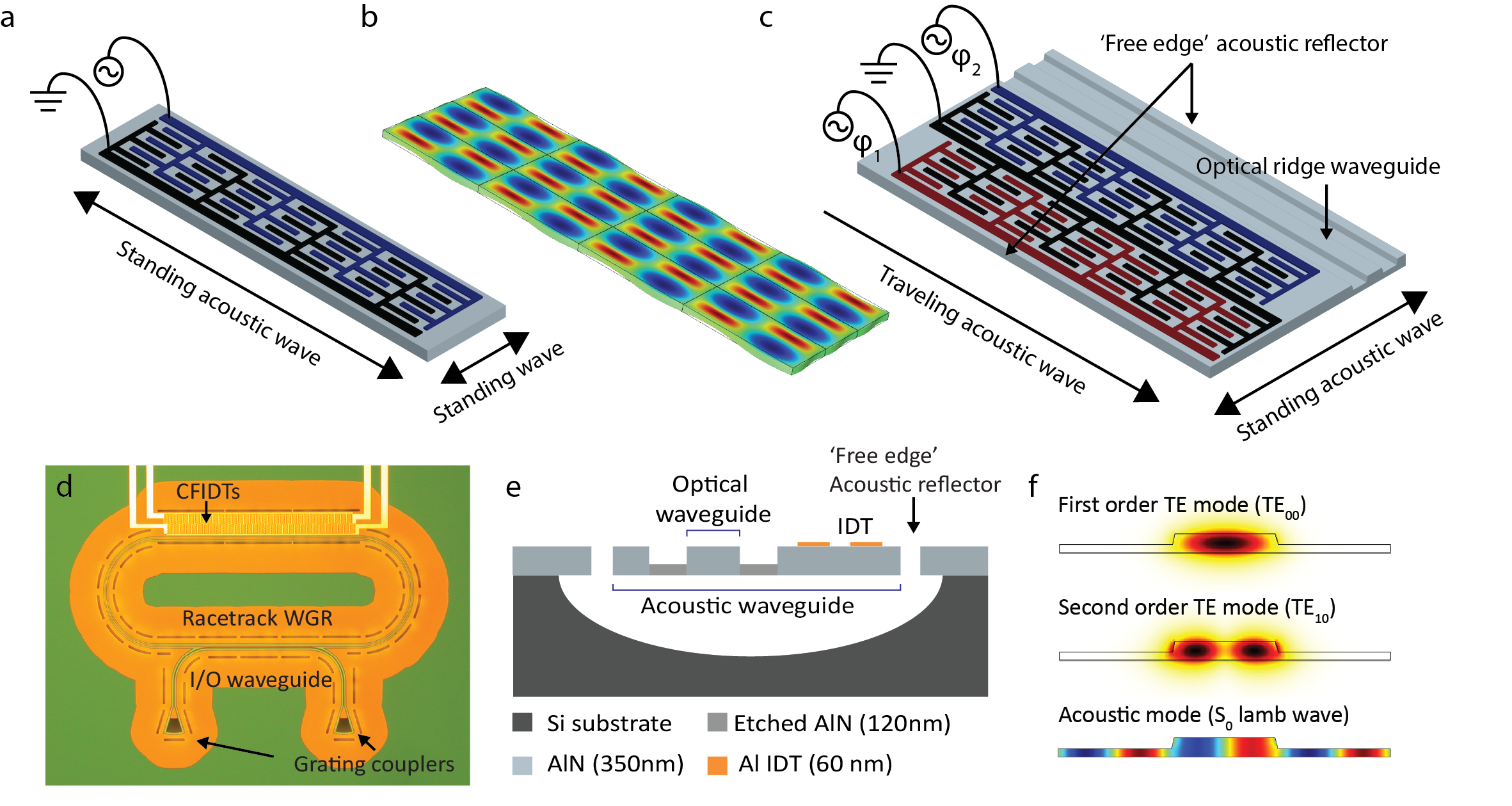}}
		\centering
		\caption{\textbf{Direction reconfigurable non-reciprocal modulator}
		\textbf{(a)} Schematic of a cross-finger interdigitated transducer (CFIDT) and
		\textbf{(b)} the corresponding S0 Lamb acoustic mode shape simulated by finite-element method. Color represents mechanical density variation with red high and blue low.
		\textbf{(c)} Schematic of the acousto-optic interaction region. Two CFIDTs are employed with quarter wavelength physical offset in the light propagating direction. Two 'free edge' acoustic reflectors are designed in the transverse direction to confine the acoustic mode. 
		\textbf{(d)} True-color microscope image of the fabricated device. Orange region is AlN on air, and green region is AlN on silicon. 
		\textbf{(e)} Cross-sectional schematic of the acousto-optic interaction region.
		\textbf{(f)} Finite-element simulated mode shapes of the TE$_{10}$ and TE$_{00}$ optical modes (electric field intensity), and the S$_0$ acoustic wave (density).  
		}
		\label{Recon2}
	\end{adjustwidth}
\end{figure}

	In order to couple two distinct optical modes in a photonic system, the photoelastic perturbation should not only satisfy the phase matching condition in the propagating direction but must also break the orthogonality in the transverse direction. 
	As shown in Fig~\ref{Recon2}f, the cross-sectional shapes of the TE$_{00}$ mode $\phi_1(x,y)$ and the TE$_{10}$ mode $\phi_2(x,y)$ respectively have symmetric and anti-symmetric electric field profile relative to the center of the waveguide. Since the optomechanical coupling $g$ is proportional to the cross-sectional overlap integral $\int\int(\phi_1\cdot\phi_2)(\nabla\cdot\psi) \, dx \, dy$ \cite{Sohn18}, the cross-sectional shape of the acoustic wave $\psi(x,y)$ should have an asymmetric density profile to ensure that the overlap integral is non-zero.
	
	To generate an acoustic wave that produces the requisite photoelastic perturbation in both propagating and transverse directions, we have developed a new cross-finger interdigitated transducer (CFIDT) illustrated in Fig.~\ref{Recon2}a. Unlike conventional IDTs, CFIDTs have electrode finger pair periodicities in two orthogonal directions. This permits electrical actuation of an acoustic wave that has a 2D profile on the propagation plane (Fig.~\ref{Recon2}b) with anti-nodal points located at the electrode fingers.
	The CFIDT pitch in the propagating direction is designed to satisfy the phase matching condition while the transverse pitch is independently designed to maximize the overlap integral between the optical modes and the acoustic wave. 
	In past work \cite{Sohn18} the propagating and transverse wave vectors were necessarily coupled since they were set by delicate tuning of the angle and pitch of a conventional IDT, which significantly complicated the design and iteration cycle.
	Finally, since the acoustic wave generated by one CFIDT alone is a standing wave, two CFIDTs are needed to produce a traveling wave as described above. In this work, we combine two CFIDTs having quarter wavelength physical offset in the direction of the optical waveguide (Fig.~\ref{Recon2}c). With an appropriate reconfiguration of temporal phase $\theta_t$, this arrangement enables generation of a propagating acoustic wave in either direction, while simultaneously forming a standing acoustic wave in the transverse direction. The transverse standing wave is supported by `free edge' acoustic reflectors on both sides of the interaction region that also resonantly enhance the acoustic wave.

	\vspace{12pt}

	\section*{Device fabrication}
	
	We fabricated the reconfigurable nonreciprocal modulator using 350 nm sputter deposited c-axis oriented aluminum nitride (AlN) on a silicon handling wafer (Fig.~\ref{Recon2}d,e). AlN is an excellent material for integrated acousto-optics due to its low intrinsic loss for both acoustics and optics, and since it has an appreciable piezoelectric coefficient.
	%
	We first pattern the two-mode racetrack resonator, along with adjacent single-mode waveguide and grating couplers using electron-beam lithography on ZEP-520 E-beam resist.	The width of the racetrack waveguide and the single-mode waveguide are selected as 2.2 \text{$\mu$}m and 0.8 \text{$\mu$}m, respectively. The waveguide is evanescently coupled to the racetrack resonator with a 1.2~\text{$\mu$}m gap over a 170 \text{$\mu$}m coupling length.
	Both the racetrack resonator and waveguide have a ridge structure formed by partially etching 230~nm of the AlN layer through Cl$_2$-based inductively coupled plasma reactive ion etch (ICP-RIE).
	Next, we pattern the `free edge' acoustic reflectors and release holes on doubled-spin coated ZEP-520 resist followed by complete etching of 350 nm AlN through Cl$_2$ based ICP-RIE.
	Two CFIDTs are then patterned by E-beam lithography followed by E-beam deposition of 60 nm of aluminum and lift off. 
	Finally, the device is released using XeF$_2$ dry etching of Si for improving the confinement of both acoustic and optical modes in the AlN waveguide. 
	
	The two CFIDTs are designed with the required quarter wavelength offset in the propagating direction along the resonator. The required acoustic wavenumber $q$, which is the wavevector difference between the two optical modes, is calculated through finite element (COMSOL) simulation based on the measured dimensions of the waveguide of the racetrack resonator. In our case, the pitch of the CFIDT in the propagating direction is set to $\Lambda_{propagating} = 2\pi/q = 18.3$~\text{$\mu$}m. The pitch in the transverse direction is set to  $\Lambda_{transverse}\,=\,2.2$~\text{$\mu$}m, which matches with the racetrack width, to maximize overlap integral.

	\section*{Measurement of non-reciprocal modulation}	
\begin{figure}[t!]
	\begin{adjustwidth}{-1in}{-1in}
		\makebox[\textwidth][c]{\includegraphics[width=1.3\textwidth]{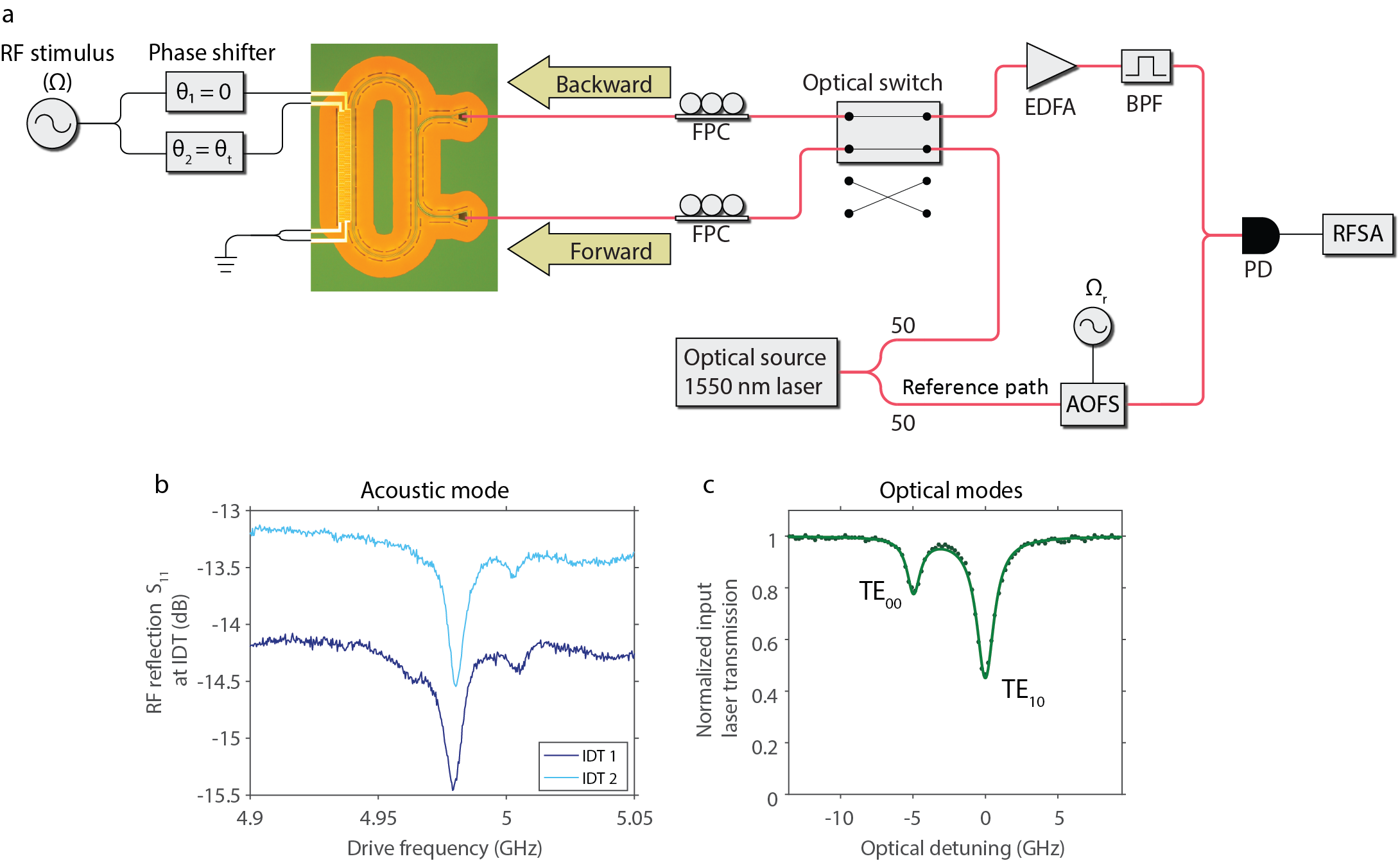}}
		\centering
		\caption{\textbf{Experimental set-up and characterization.}
		\textbf{(a)}
		Light from a continuous 1520 - 1570 nm fiber-coupled laser is split into two branches for optical heterodyne detection. The light in the reference path is offset by $\Omega_r$ = 100 MHz using an acousto-optic frequency shifter (AOFS) to enable separate measurement of Stokes and anti-Stokes sidebands. 
		Fiber polarization controllers (FPCs) are used to manipulate the polarization of the input and the output, to orient correctly with the on-chip grating couplers. 
		The output light from the device is amplified using an erbium-doped fiber amplifier (EDFA) to compensate for the loss from the grating couplers.
		A tunable optical band pass filter is placed to minimize additional broad spectrum noise added by the EDFA. 
		The light from the reference path and the device is combined and generates beat notes at a photodetector (PD). The RF signal from the PD is measured using a RF spectrum analyzer (RFSA).
		An optical switch is used to control the direction of light going into the device.
		The RF stimulus signal -- which controls the non-reciprocity -- is produced by a signal generator and split into two paths connected to each CFIDT. Phase shifters are placed at each path to independently control the phase of the two RF inputs to the CFIDTs.
		\textbf{(b)}
		Measured RF reflection coefficients (S$_{11}$) of the two CFIDTs obtained with a vector network analyzer. Both CFIDTs efficiently generate the acoustic excitation at 4.98 GHz, however, do have slightly different electromechanical transduction efficiency. 
		\textbf{(c)}
		Measured (dots) and fitted (solid) optical transmission spectrum through the waveguide shows the optical mode pair with the frequency separation of 4.97 GHz located around 1540 nm. The left dip is the resonance of the TE$_{00}$ mode and the right dip is the TE$_{10}$ mode. 
		}
		\label{Recon3}
	\end{adjustwidth}
\end{figure}

	We begin experiments by first characterizing the acoustic and optical properties of the system. The CFIDTs are characterized by means of RF reflection ($s_{11}$) measurement using a vector network analyzer (VNA). As shown in Fig.~\ref{Recon3}b, both CFIDTs exhibit the same resonant frequency for the S0 Lamb acoustic mode located at 4.98 GHz. The optical transmission spectrum through the waveguide allows measurement of the optical modes of the resonator (Fig.~\ref{Recon3}c). Here, we select a resonator mode pair located near 1540~nm with frequency separation of 4.97 GHz, which is very similar to the acoustic resonance frequency. The measured optical transmission spectrum (Fig.~\ref{Recon3}c) is plotted relative to the TE$_{10}$ resonance frequency ($\Delta=\omega_l-\omega_1=0$). The loaded quality factor of the TE$_{00}$ mode is measured at $\sim$176,000 and of the TE$_{10}$ mode is $\sim$133,000. Both of the resonator modes are under-coupled to the single-mode waveguide.

	We experimentally test non-reciprocal modulation by measuring the Stokes and anti-Stokes optical sidebands produced independently for forward and backward optical inputs, for which optical heterodyne measurement (Fig.~\ref{Recon3}a) is employed. For this, we first produce a 100 MHz frequency-shifted optical reference from the input laser signal using an acousto-optic frequency shifter (AOFS). This photocurrent beat notes formed between this reference signal and either the Stokes or anti-Stokes sidebands of the original input are therefore frequency-separated and can be measured independently.
	The electrode drive stimulus from a signal generator is split into two paths -- the input to one CFIDT is provided through an RF phase shifter which allows control of the temporal phase $\theta_t$, while the input to the second CFIDT is unmodified. 
	%
	During this experiment the RF drive frequency is set at 4.97~GHz so that the acoustic wave is most efficiently generated for both CFIDTs, and the RF input power provided to each CFIDT is 2 dBm.

\begin{figure}[t!]
	\begin{adjustwidth}{-1in}{-1in}
		\makebox[\textwidth][c]{\includegraphics[width=1.3\textwidth]{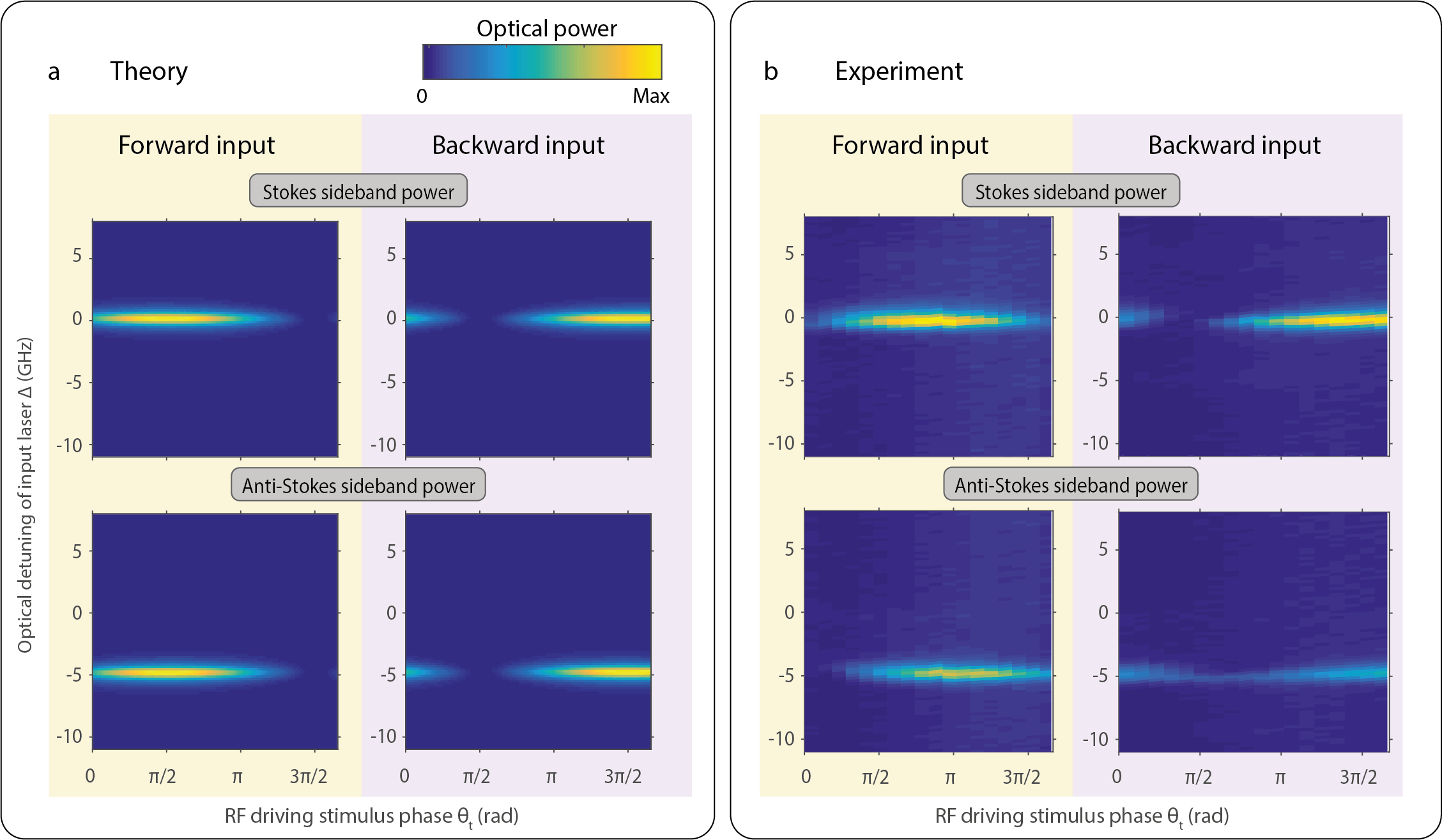}}
		\centering
		\caption{\textbf{Experimental demonstration of direction reconfigurable non-reciprocal modulation.} 
		\textbf{(a)} Theoretical Stokes and anti-Stokes sideband power with respect to the input laser optical detuning ($\Delta$) and RF driving stimulus phase difference ($\theta_t$). The plot is based on the parameters extracted from carrier transmission data (Fig 3c) and the predictive Eqns. 5.
		Note that the detuning $\Delta$ is defined relative to the higher frequency, i.e. the TE$_{10}$, optical mode in this experiment. As a result, $\Delta=0$ implies that the laser is on resonance with the higher frequency mode, and that only a Stokes sideband should be observable when scattering takes place. Similarly, $\Delta = -5$ GHz implies that the laser is on resonance with the lower frequency mode and that only the anti-Stokes sideband should be observed.
		\textbf{(b)} Measured Stokes and anti-Stokes data in the forward and backward direction shows the nonreciprocal modulation. 
		}
		\label{Recon4}
	\end{adjustwidth}
\end{figure}

	In Fig.~\ref{Recon4}a, we present a theoretical prediction of the output power of Stokes and anti-Stokes sidebands as a function of detuning of the input laser $\Delta$ and the temporal phase applied on RF stimulus $\theta_t$. 
	In Fig.~\ref{Recon4}b we present the experimentally measured Stokes and anti-Stokes sidebands that exhibit the non-reciprocal modulation.

	%

	Let us first examine intermodal scattering at $\theta_t=0$ where a standing acoustic wave is generated. This corresponds to the left-most edge of each of the subplots in Fig.~\ref{Recon4}. Since  the acoustic excitation is non-propagating, the intermodal scattering occurs in both forward and backward directions symmetrically.
	When the light enters the TE$_{10}$ mode (at $\Delta = 0$ GHz) in either direction, only a single Stokes sideband is generated. This is because Stokes scattering from the TE$_{10}$ mode is resonantly enhanced by the TE$_{00}$ resonance located at $\Delta= -4.97$~GHz. However, anti-Stokes scattering is suppressed since there is no optical mode that is phase matched. Similarly, if the light enters the TE$_{00}$ mode, only the anti-Stokes sideband is generated due to the TE$_{10}$ resonance located at $\Delta= 0$. Using this data, we measured a maximum modulation efficiency (output sideband power vs. input optical power) of 0.0025\% when the system is driven with 5~dBm of RF input power.

	We now test for reconfiguration of this non-reciprocal behavior by analyzing the sideband response as a function of the temporal phase difference $\theta_t$. We focus on the specific cases $\theta_t= 0, \pi/2, \pi, 3\pi/2$ where the acoustic excitations have properties of interest: at $\theta_t= 0$ and $\pi$ the excitation is a purely standing acoustic wave, at $\theta_t=\pi/2$ the excitation is a pure backward propagating acoustic wave, and at $\theta_t=3\pi/2$ the excitation is a pure forward propagating acoustic wave. 
	%

	In both experimental data and theoretical simulation, there is a clear distinction in the sidebands produced depending on whether the light enters in the backward or forward direction. In the theoretical plot, when $\theta_t = 0$ and $\pi$, the sideband amplitudes in the backward and forward directions are identical. When the phase difference is set to $\theta_t=\pi/2$, the backward sideband generation is suppressed while the sideband amplitude in the forward direction is maximized. 
	This is exactly as expected since $\theta_t=\pi/2$ only launches an acoustic wave in backward direction, implying that only the forward optical modes are correctly phase matched.
	Conversely, when $\theta_t=3\pi/2$, the forward sideband generation is suppressed while scattering in the backward direction is maximized.

	The experimental data shows very similar trends. However, the sideband amplitude peaks and nulls are slightly misaligned on the $\theta_t$ axis relative to the theoretical prediction. For example, the peak of anti-Stokes sideband power in forward direction is located $\theta_t \approx 3\pi/4$, whereas the null of the anti-Stokes sideband power in the backward direction is located near $\theta_t = \pi/2$. 
	We expect that this shift is caused by a small quantity of undesired scattering (i.e. intramodal scattering), which can interfere inside the single mode waveguide with the desirable non-reciprocal intermodal scattering, thereby shifting the apparent null points. Additional factors that contribute to this shift may be slight physical misalignment of the two CFIDTs, or the physical phases of the acoustic waves produced by them, either of which may be due to fabrication imperfection. Resolution of this question is not possible with the capabilities of the current devices and is left for future work.

	The largest sideband non-reciprocity contrast that we experimentally observed is $\sim$8 dB near $\theta_t = \pi/2$, and is also primarily limited by the non-ideality of the experiment. Specifically, if the amplitudes of the two acoustic waves generated by the CFIDTs are slightly different, the synthesized acoustic wave cannot be a pure traveling wave and does contain a standing wave term. This would creates a little extra scattering that might be possible to calibrate out by setting different drive amplitudes at the two CFIDTs.
	In practice, this inequality is readily caused by differences in electromechanical couplings of the CFIDTs as shown in the $s_{11}$  measurement (Fig.~\ref{Recon3}b), or from different RF losses in the transmission lines connecting the RF power splitter to the CFIDTs.

\vspace{24pt}

\section*{Conclusion}

In this work, we have demonstrated a dynamically reconfigurable non-reciprocal acousto-optic modulator at telecom wavelength, which can be set in a variety of reciprocal and non-reciprocal configurations. The non-reciprocal modulation is achieved by biasing the optical resonator with a synthesized traveling acoustic wave, with the direction and magnitude of non-reciprocity simply controlled by adjusting the temporal phase of an RF stimulus. This reconfigurable functionality can potentially enable new signal processing capabilities for integrated communications and sensing systems. Moreover, this work showcases a fully lithographically-defined approach to producing non-reciprocal components, that is not strictly material dependent, and that can be adapted for different wavelength regimes. We expect that the performance of this non-reciprocal acousto-optic modulator can be further improved by enhancing electro-mechanical coupling, acousto-optic coupling, and optical quality factors of the sub-components. In the long term, this type of approach could be adapted to produce linear optical isolators and circulators in integrated photonics.

\vspace{24pt}

\newpage

\bibliographystyle{naturemag}
\bibliography{recon2}
\newpage

\end{document}